# Sorption of $^4$He, $H_2$, Ne, $N_2$, $CH_4$ and Kr impurities in graphene oxide at low temperatures. Quantum effects.

A.V. Dolbin[1], V.B. Esel'son[1], V.G. Gavrilko[1], V.G. Manzhelii[1], N.A.Vinnikov[1], R.M. Basnukaeva[1], V.V. Danchuk[1], N.S. Mysko[1], E.V. Bulakh[2], W.K. Maser[3] and A.M. Benito[3]

[1]B. Verkin Institute for Low Temperature Physics and Engineering of the National Academy of Sciences of Ukraine, 47 Lenin Ave., Kharkov 61103, Ukraine.

[2] Institute of Radio Astronomy (IRA) of the National Academy of Sciences of Ukraine, 4 Krasnoznamennaia Str., Kharkov 61002, Ukraine.

[3] Instituto de Carboquímica ICB-CSIC, 4 Miguel Luesma Castán, E-50018 Zaragoza, Spain.

Electronic address: dolbin@ilt.kharkov.ua

PACS:
68.43.Jk Diffusion of adsorbates, kinetics of coarsening and aggregation

**Abstract**

Sorption and the subsequent desorption of $^4$He, $H_2$, Ne, $N_2$, $CH_4$ and Kr gas impurities by graphene oxide (GO), glucose-reduced GO (RGO-Gl) and hydrazine-reduced GO (RGO-Hz) powders have been investigated in the temperature interval 2-290 K. It has been found that the sorptive capacity of the reduced sample RGO-Hz is three to six times higher than that of GO. The reduction of GO with glucose has only a slight effect on its sorptive properties. The temperature dependences of the diffusion coefficients of the GO, RGO-Gl and RGO-Hz samples have been obtained using the measured characteristic times of sorption. It is assumed that the temperature dependences of the diffusion coefficients are determined by the competition of the thermally activated and tunneling mechanisms, the tunneling contribution being dominant at low temperatures.

**Keywords:**

Graphene oxide, glucose-reduced graphene oxide, hydrazine-reduced graphene oxide, gas impurities, sorption, desorption, diffusion coefficients, tunneling.

## 1. Introduction

Graphene is a one-layer two-dimensional carbon structure whose surface consist of regular hexagons with a 1,42 Å side [1] and $sp^2$-hybridized carbon atoms at each vertex. A similar structure is a constituent of crystalline graphite in which such graphene planes are ~3.35 Å apart from each other [1]. The recent discovery of a fairly simple technique of extracting individual graphene and obtaining macroscopic quantities of graphene-based materials, such as graphene oxide have stimulated keen interest in production, investigation and practical use of graphene. The interest is primarily related to the unique physical and chemical properties of graphene – its high electric [2] and thermal [3] conductivities the dependence of the electron characteristics upon various – origin bound radicals at the graphene surface [4], the controllable width of the forbidden band [5], the quantum Hall effect [6], high elasticity, remarkably high electron mobility [7], and outstanding electric and mechanical characteristics in the MHz-range [8]. Owing to these properties, graphene shows considerable promise as a basis for developing new nanomaterials with improved mechanical, electric and thermophysical properties. It can also be used as high-efficiently gas [9, 10] and biological [11, 12] sensors.

Oxidation of graphite and the subsequent ultrasonic dispersion of the graphite oxide to the state of graphene oxide (GO) is one of the most effective technique towards large-scale production of graphene-based materials [13, 14]. Graphene oxide consists of intact graphite areas with inclusions of $sp^3$-hybridized carbon atoms. The inclusions contain hydroxyl and epoxy functional groups at the upper and bottom surfaces of every graphene sheet and $sp^2$-hybridized carbon containing carboxyl and carbonyl groups located mainly at the edges of the graphene sheet.

Graphene oxide, like graphite, has a layered structure consisting normally of a few graphene layers. Typically its number is about 2 to 10 layers, depending on the exfoliation degree achieved. The layer spacing vary within 6-8 Å depending on the technology of GO preparation and the oxidation level [15, 16]. The carbon layers of GO are deformed by the $sp^2 \rightarrow sp^3$ transitions of the C atoms. GO is capable of binding ions of some metals in solutions and interacting with organic and inorganic compounds. As a result, it is possible to obtain porous carbon materials containing particles of metals (Pt, Pd, Au, etc) [17, 18]. Normally, GO contains a great number of topological structural defects and ruptures. The GO layers are weakly related to each other.

The $O_2$ – containing groups can be removed partially by reducing graphene oxide, for example, with hydrazine, dimethyl hydrazine, hydroquinone, sodium boron hydride and so on [19]. It is however essential that the residual oxide groups and the surface defects, inevitable in the process of GO reduction, affect seriously the structure of the graphene plane and a complete recovery of the $sp^2$ carbon structure is highly unlikely. The RGO technology is one of the methods used to produce macroscopic quantities of a graphene-like material for wide-range applications, such as electrochemical energy storage, sensing, composites and catalysis [20]. Medical applications (for example, for target-oriented medicine delivery [21]) require RGO free even of traces of toxic reducing agents, such as hydrazine. This can be achieved by using "green" methods of reducing GO with glucose [22] and other non-toxic substances [23, 24].

Derivatives of GO and RGO possess large specific surface areas [25, 26] and are often used as high-efficiency sorbents. It is therefore important to investigate the sorptive properties of GO and RGO in a wide temperature interval for different types of impurities. Besides, the periodicity of the potential at the cellular structure of graphene suggests that at low temperatures the tunnel motion of light impurity particles is quite probable both at the surface of single graphene planes (see theoretical studies [27, 28]) and between the neighboring planes of several GO and RGO layers. Earlier, we observed tunnel diffusion of He isotopes, $H_2$ and Ne in experiments on their saturation/desorption in samples of fullerite $C_{60}$ [29, 30] and bundles of single-walled carbon nanotubes [31]. At low temperatures the tunnel motion of He isotopes in bundles of carbon nanotubes caused high negative thermal expansion [32, 33]. No experimental information about this sort of effects has been reported for graphene oxide.

## 2. Experimental technique

The sorptive properties and the saturation-desorption kinetics of $^4$He, $H_2$, Ne, $N_2$, $CH_4$ and Kr gas impurities in graphene oxide (GO) and hydrazine – reduced graphene oxide (RGO-Hz) have been investigated experimentally using a laboratory stand (its design and operation are detailed in [29, 34]). The temperature interval was 2-290 K. The GO and RGO-Hz samples were powders with masses of ~40 mg.

Graphite Oxide was prepared using Hummers' method from graphite powder (Sigma-Aldrich) by oxidation with $NaNO_3$, $H_2SO_4$, and $KMnO_4$ in an ice bath as reported elsewhere [13]. A suspension of graphene oxide (GO) sheets was obtained by sonication of the prepared graphite oxide powder in distilled water (1 mg/mL) for 2 h, followed by mild centrifugation at 4500 rpm for 60 min to remove non-exfoliated materials, leading to a brown-colored dispersion of exfoliated GO sheets. GO dispersion was freeze-dried to obtain GO powder material [35].

Two different reductive techniques were applied to the GO dispersion to obtain the corresponding reduced graphene oxide materials: i) addition of an excess of hydrazine hydrate $N_2H_4 \cdot H_2O$ (6 µL/mL GO dispersion) and refluxing for 5 hours (RGO-Hz); and ii) addition of glucose (6.4 mg/mL GO dispersion) and refluxing for 3 hours (RGO-Gl). The corresponding RGO powders were obtained by filtration of the reduced GO dispersions through a polycarbonate membrane filter of 3 µm pore size, followed by washing with 200 mL of distilled water and vacuum drying at 80 °C for 48 h.

The morphology of the GO, RGO-Hz and RGO-Gl samples was investigated by optical microscopy (MBS-10 microscope with a CCD digital camera) and transmission electron microscopy (EM-125K TEM). In the optical photographs the samples are macroscopic agglomerates of mainly parallel-packed graphene planes (Fig.1a,b). Note that although the original GO dispersed material consisted of individual few layered flakes, depositions of non-highly diluted dispersions on surfaces results in agglomerations. Also, the drying process to obtain GO powder material favors agglomeration of individual GO flakes. Therefore, these agglomerated forms are the morphologies to be considered in the sorption studies. The optical photographs were analyzed using a program package for microscopic image processing (ImageJ) and the observed conglomerates were graded according to their areas and lateral sizes. In reduced GO 4000 μm$^2$ –20000 μm$^2$ conglomerates (grains) are dominant and their largest area is up to 70000 μm$^2$. GO consists mainly of small 250 μm$^2$ – 5000 μm$^2$ conglomerates though formations with areas over 200000 μm$^2$ are also available. It is obvious that the process of reducing GO provokes destruction of large conglomerates and the subsequent formation of agglomerates with close-size grains. Proceeding from the above analysis, we estimated the average area and the characteristic average size of the grain (~150 μm for GO and ~120 μm for reduced GO). The areas of agglomerates consisting of several stuck-together grains were disregarded. Note that the obtained characteristic grain sizes are rather approximate estimates since the sizes and the configurations of carbon planes making up individual grains can vary considerably (typically individual GO and RGO flakes are about 1 to 10 µm in lateral size [36–39]).

The performed TEM investigations allowed us to observe the morphology of single planes in both GO and RGO (Fig.1c,d). While in the parent GO material individual flakes are hard to observe, upon the reduction, RGO individual flakes can be easily seen although overlapping each other thus forming the base for the larger agglomerates as observed in b).

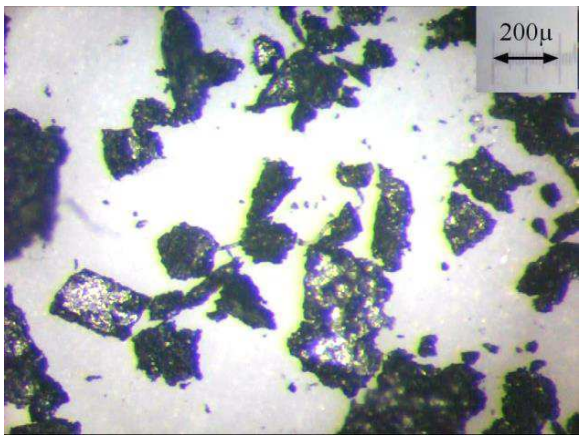 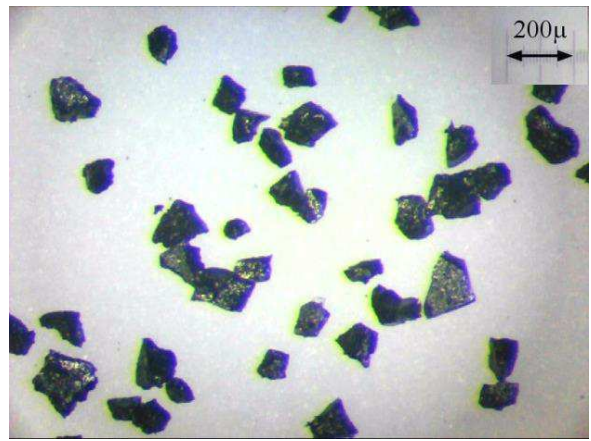

a)  b)

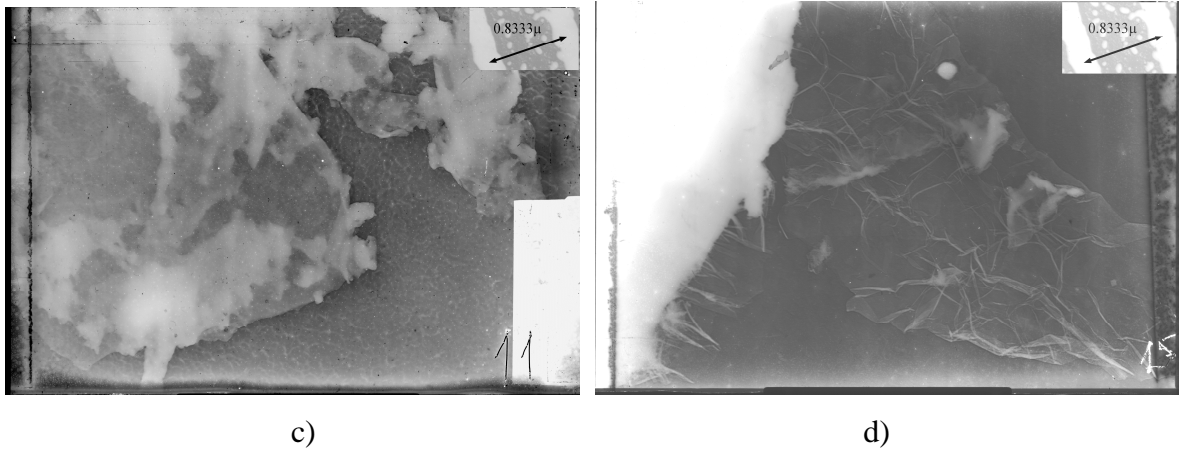

|  |  |
|:---:|:---:|
| c) | d) |

Fig.1. Optical photographs (a,b) and TEM images of GO and RGO samples: a) and c) GO; b) and d) RGO-Gl.

Immediately before starting the investigation, the powder samples were evacuated for five days at room temperature in the measuring cell of the stand to remove the possibly available gas impurities and moisture.

Graphene oxide and the reduced materials were saturated with used impurities under a pressure about 1 Torr at the lowest possible temperature for each particular impurity. As soon as the sample absorbed the gas, a new portion of impurity was added to the cell. The saturation temperature was chosen so that the pressure of the gas impurity in the measuring cell remained in the process of saturation 2.5 – 3 times lower than the saturation vapor pressure of the impurity at this temperature. These sorption condition excluded condensation of the impurity vapors and the formation of films at the grain surfaces and the cell walls. The gas impurity flow was stopped when the equilibrium pressure $10^{-2}$ Torr was achieved in the cell. The cell was then sealed and the changes in the pressure were recorded in the process of impurity desorption from the powder on its stepwise heating. The gas impurity evolved due to the heating was taken to an evacuated calibrated vessel. The pressure in it was measured with two capacitance manometer (MKS-627B) capable of measuring pressure in the range $10^{-3}$ – 1000 Torr, the error being $1 \cdot 10^{-4}$ Torr. When the pressure was stabilized, the vessel was detached from the sample, evacuated and then joined to it again. The gas extraction was continued until the gas pressure over the sample reduced to $10^{-2}$ Torr. Then the measurement was repeated at the next value of temperature. When the impurity was completely removed at room temperature, the sample was cooled again down to 2 K and the saturation-desorption process was repeated for the next impurity.

The experimental time dependences of the pressure variations ΔP during sorption (or desorption) of the $^4$He, $H_2$, Ne, $N_2$, $CH_4$ and Kr impurities by the GO and RGO powders are described well by the exponential function (Fig.2):

$$\Delta P = A \cdot (1 - \exp(-t/\tau)) \quad , (1)$$

where the parameter $\tau$ corresponds to the characteristic time during which the impurity particles occupy the grains of the GO powder, $A$ is the fitting parameter.

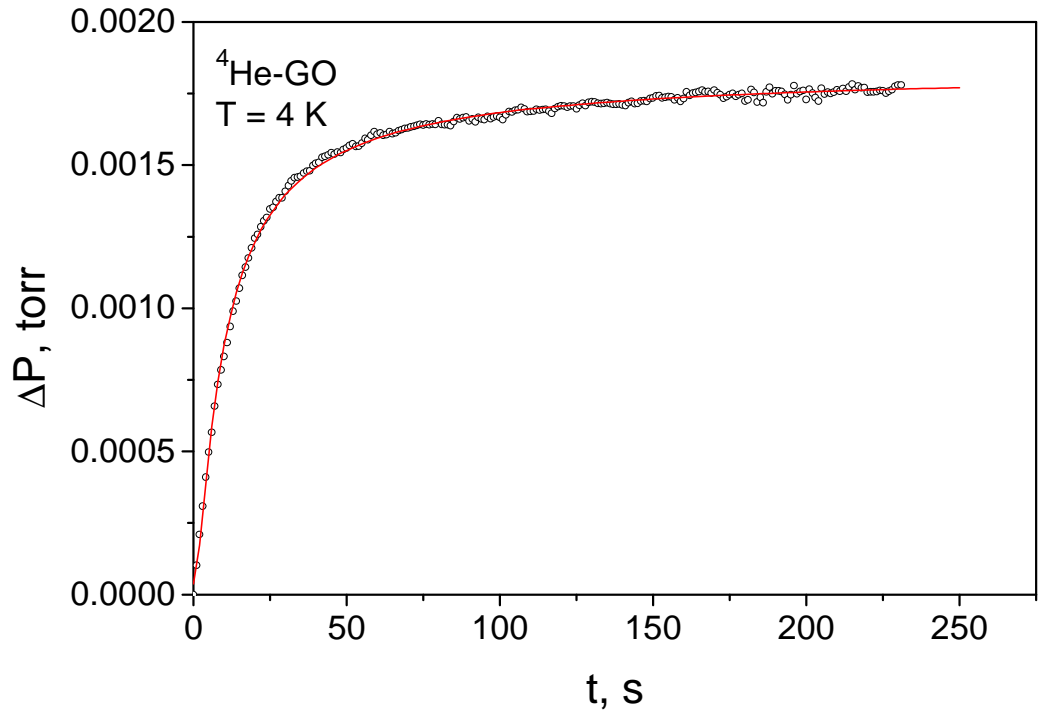

Fig.2. Experimental results on pressure variations in the process of $^4$He desorption from the GO sample (circles) and their approximations by Eq.(1) (solid line, $\tau \sim 9$ s). The temperature of the sample is T=4K.

Note that the pressure vs time dependences measured on sorption-desorption of impurities at the same temperature of the sample were similar qualitatively and their characteristic times coincided within the experimental error. The characteristic times $\tau$ of sorption/desorption were used to estimate the diffusion coefficients of the impurities in GO and RGOs:

$$D \approx \frac{\overline{\ell}^2}{4 \cdot \tau} , \qquad (2)$$

where $\overline{\ell}$ is average grain size of the GO and RGO powders, $\tau$ is the characteristic time of diffusion.

Since the impurity diffusion proceeded mainly along the graphene planes, the proportionality coefficient in the denominator of Eq.(2) for the case close to two-dimensional diffusion is equal to 4.

## 3. Results and discussion

The temperature dependences of the gas quantities desorbed from GO, RGO-Hz and RGO-Gl samples are shown in Fig.3. The total quantity of the desorbed gas impurities is given in Table 1. $N_A$ in Fig.3 and Table 1 is the quantity normalized to the total number of C atoms $N_C$ in the samples. Note that the total quantities of the sorbed and desorbed impurities coincide within the experimental error.

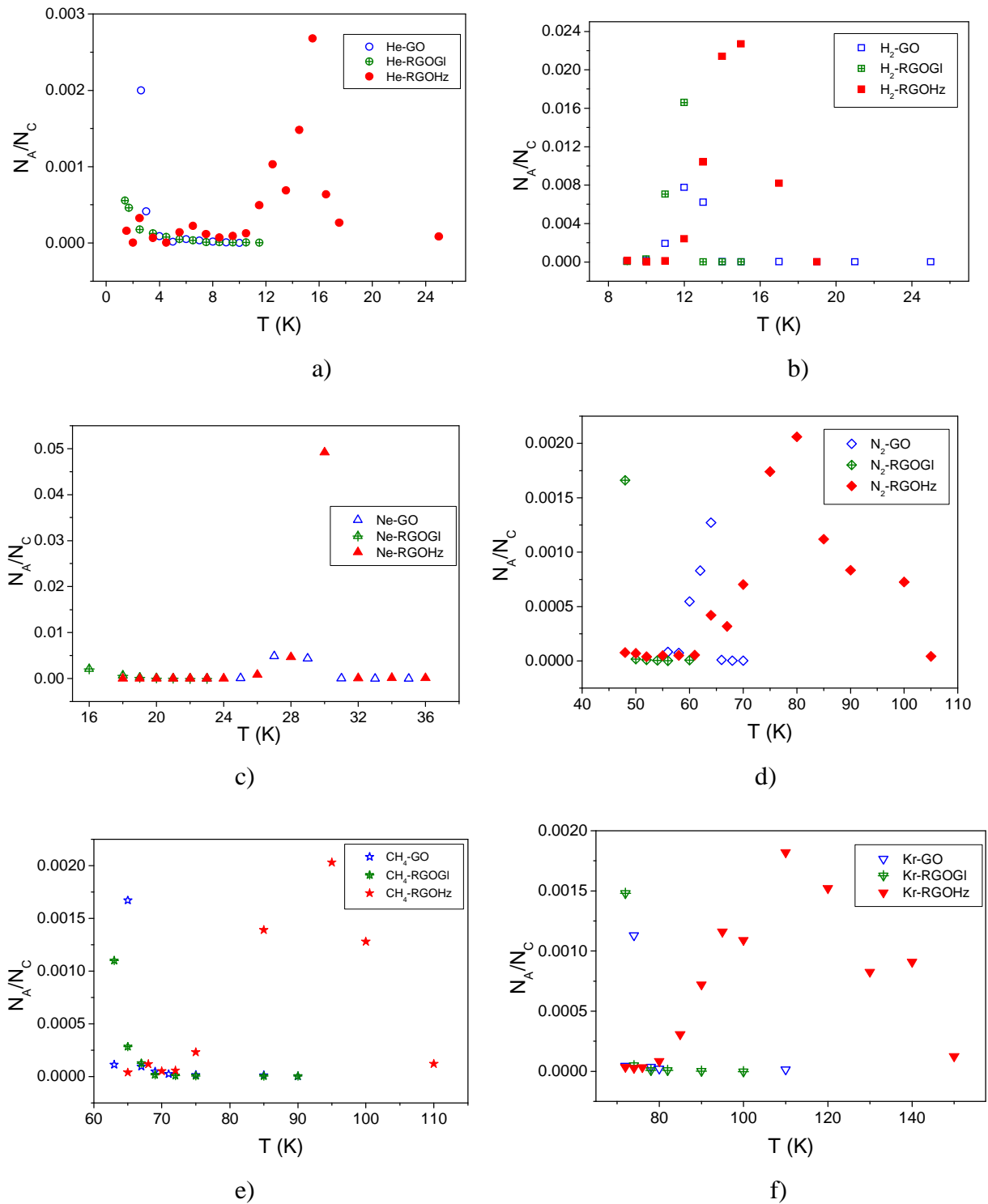

Fig.3. The temperature dependences of the relative quantity of the impurity desorbed from the GO, RGO-Hz and RGO-Gl samples; a) $^4$He, b) $H_2$, c) Ne, d) $N_2$, e) $CH_4$, f) Kr.

Table 1. The total concentrations of the impurity particles in the GO, RGO-Gl and RGO-Hz samples (in molar fractions, i.e. the number of impurity particles per C atom).

| A, impurity | $^4$He | $H_2$ | Ne | Kr | $N_2$ | $CH_4$ |
|---|---|---|---|---|---|---|
| $N_A/N_C$ (GO) | 0.0026 | 0.016 (0.27%wt) | 0.0079 | 0.0013 | 0.0028 | 0.00198 |
| $N_A/N_C$ (RGO-GL) | 0.00151 | 0.024 (0.4%wt) | 0.0012 | 0.0016 | 0.0017 | 0.00156 |
| $N_A/N_C$ (RGO-Hz) | 0.00867 | 0.0654 (1.1%wt) | 0.0551 | 0.00866 | 0.00829 | 0.00531 |

It is seen in Fig.3 that the temperature intervals of the dominant bulk of impurities desorbed are different for the GO, RGO-Gl and RGO-Hz samples. This may be due to the features of the sample structures. It is probable that the access of the impurity particles to the interplanar space of GO is limited by the oxide groups. As a result, the sorption of impurities proceeds mainly at the outer surfaces of the graphene planes. The Gl-reduction of GO had little effect on the desorption curves. This may be due to adsorption of the reducing molecules (glucose in our case) (cf. [41]) which, much like oxide groups, block the layer spacings in RGO-Gl and thus increase the interplanar distance (up to 1 – 1.4 nm [22, 35]). On the contrary, the Hz-reduction decreases the layer spacing (to 0.55 nm [22] or even to 0.44 nm when epoxy and hydroxyl groups are almost completely removed at both sides of the graphene plane [42]). The Hz-reduction of GO changes drastically the character of the desorption curves by shifting the desorption maximum towards higher temperatures for all of the investigated impurities. Most likely the Hz-reduction of GO is an efficient way of removing the functional groups which block the access to the space between the graphene planes. Since the impurity particles in the interplanar space of the RGO-Hz sample have higher binding energies than those at the outer graphene surface [43], the particle desorption from the interplanar space proceeds at a higher temperature. The assumption of a greater possibility for sorption in the interplanar space of the RGO-Hz sample is supported by the higher (3-6 times) sorptive capacity of this sample in comparison with the GO and RGO-Gl samples (see Table 1). Our values of the capacity of $H_2$ sorption in Hz-reduced GO agree well with literature data (1.2 wt% and 0.1 wt% at P~10 bar, T=77 and T=298 K respectively [44]; 1.7 wt% at P=1 atm and T=77 K [45]). The theoretical limit of physical sorption of $H_2$ by single-layer graphene at P=1 atm and T=77K is estimated to be 3 wt% [44].

The temperature dependences of the diffusion coefficients of the $^4$He, $H_2$, Ne, $N_2$, $CH_4$ and Kr in GO and reduced GOs (RGO-Hz and RGO-Gl) are shown in Fig.4.

The temperature dependences of the diffusion coefficients have several interesting features. At the lowest temperature of the experiment the coefficients of He, $H_2$ and Ne diffusion are only slightly dependent on temperature (Fig.4a). The reason may be that at low temperatures diffusion of light impurities consisting of small-diameter molecules proceeds predominantly by the tunneling mechanism while thermally activated diffusion prevails at high temperatures. This effect is considerably weaker for heavier impurities ($N_2$, $CH_4$, Kr) and shows up as a change in the slope of the temperature dependence of the diffusion coefficients (Fig.4b).

To estimate the activation energy of impurity diffusion RGO-Hz, the temperature dependences of the diffusion coefficients (Fig.4) was plotted in the coordinates $Y=ln(D)$ vs. $X=1/T$ (see Fig.5). The dependence $Y(X)$ is linear if the process of diffusion obeys the Arrhenius equation (Eq.3)

$$D = D_0 \exp\left(-\frac{E_a}{k_B T}\right), \qquad (3)$$

where $E_a$ is the activation energy of diffusion, $D_0$ is the entropy factor dependent on the frequency of collisions between the host and impurity molecules, $k_B$ is the Boltzmann constant.

The dependence $ln\,D$ vs. $(1/T)$ has two distinct linear parts describing the diffusion of $^4$He atoms in the RGO-Hz sample (Fig.5). The slanting linear part corresponds to the temperature interval in which thermally activated diffusion is dominant. The part that runs almost parallel to the X-axis refers to the temperature interval in which the tunnel contribution to diffusion predominates.

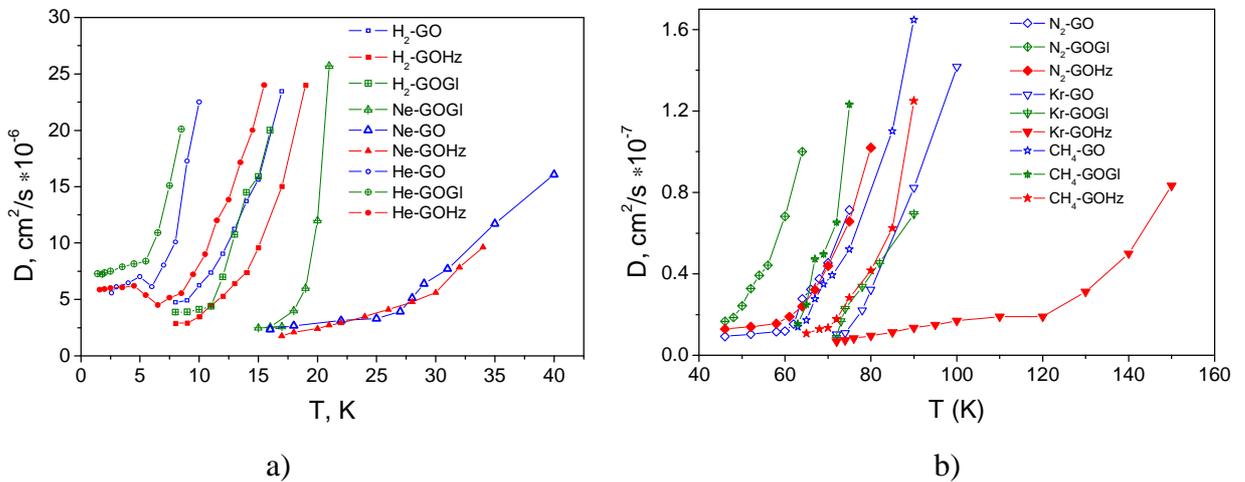

Fig.4. The temperature dependences of the diffusion coefficients of the $^4$He, $H_2$, Ne, $N_2$, $CH_4$ and Kr impurities in the GO, RGO-Hz and RGO-Gl samples: a) $^4$He, $H_2$, Ne; b) $N_2$, $CH_4$, Kr.

The obtained dependences (Fig.4, Fig5) were employed to estimate the activation energies of $^4$He, $H_2$, Ne, $N_2$, $CH_4$, Kr diffusion in the GO, RGO-Hz and RGO-Gl samples (see Table 2). We used only the diffusion coefficients belonging to the temperature interval with the prevailing thermally activated mechanism. Equation (3) is unsuitable when the contribution of quantum effects to diffusion increases.

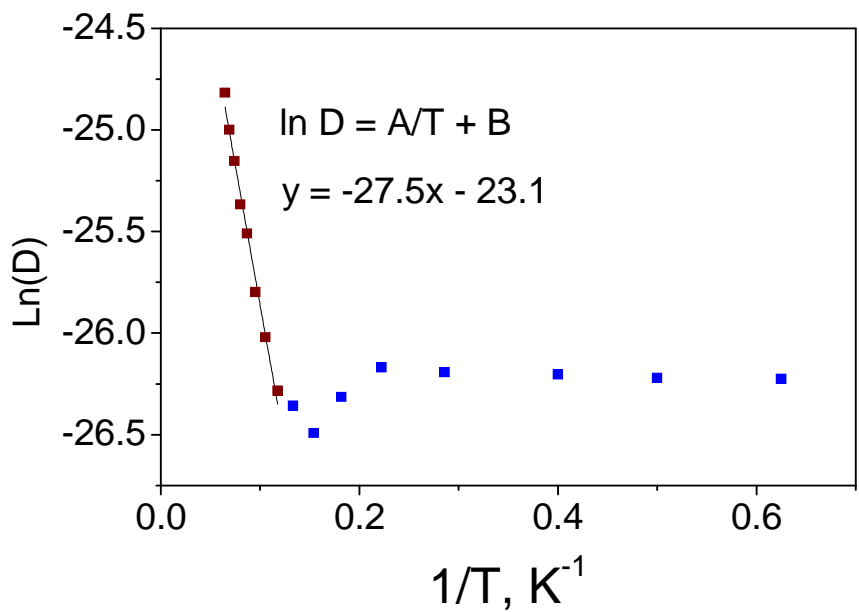

Fig.5. The $Y=ln(D)$ vs. $X=1/T$ dependence for the coefficient of $^4$He diffusion in hydrazine-reduced GO (RGO-Hz).

Table 2. The activation energies for the impurity particles in the GO, RGO-Hz and RGO-Gl samples.

| $A$, impurity | $^4$He | $H_2$ | Ne | $N_2$ | $CH_4$ | Kr |
|---|---|---|---|---|---|---|
| Effective diameter $\sigma$, Å [46] | 2.62 | 2.96 | 2.788 | 3.708 | 3.817 | 3.921 |
| Mass $m$, a.m.u. | 4 | 2 | 20 | 28 | 16 | 83.798 |
| $E_A$ (GO), K | 12.0 | 35.4 | 110.4 | 363.3 | 498.0 | 595.8 |
| $E_A$ (RGO-GL), K | 13.6 | 40.7 | 123.3 | 365.8 | 594.0 | 622.1 |
| $E_A$ (RGO-Hz), K | 27.6 | 61.6 | 138.9 | 460.1 | 659.6 | 878.5 |

It is seen in Table 2 that the activation energies of impurity diffusion in GO and its reduced modifications increase as the mass and the effective diameter of the impurity particles grow.

The highest activation energies of impurity diffusion were observed in the RGO-Hz sample. This correlated well with the expected possibility for the impurity particles to penetrate into the layer spacings of Hz-reduced GO. It is known that for light impurities ($^4$He, $H_2$) the activation energy of diffusion in a layer spacing is almost twice as high as that at the surface of graphene [43].

The obtained activation energies of $H_2$ diffusion are at least half as high as their theoretical estimates for $H_2$ physiosorbed on a graphene sheet ($E_a(H_2)<10$ MeV [47]) and several time higher than the estimated energy of the drift of a hydrogen molecule along the graphene sheet ($E_{drift} < 1$ meV [48]). These distinctions may be attributed to the structural features and numerous defects existing in real GO samples.

### 4. Conclusions

Sorption and the subsequent desorption of $^4$He, $H_2$, Ne, $N_2$, $CH_4$ and Kr gas impurities in a graphene oxide (GO) powder and its modification reduced with glucose (RGO-Gl) and hydrazine (RGO-Hz) have been investigated in temperature interval 2-290 K. The highest effect upon the sorptive properties of the samples was observed in Hz-reduced GO: the total sorptive capacity of RGO-Hz was three to six times higher than that of GO. It is possible that the $O_2$-containing groups were removed in the process of Hz-reduction, which opened the layer spacings for sorption. This assumption is supported by the increased activation energies of RGO-Hz in comparison with GO and RGO-Gl. The Gl-reduction of GO had a little effect on the sorptive properties of the RGO-Gl sample.

The measured characteristic times of saturating the GO, RGO-Gl and RGO-Hz samples with impurity particles were used to obtain the temperature dependences of the diffusion coefficients for the impurities specified. It is assumed that the behavior of the temperature dependences of the diffusion coefficients for the investigated impurities in the GO, RGO-Gl and RGO-Hz samples is determined by the competition of the thermally activated and the tunneling mechanisms of diffusion, the contribution of the latter mechanisms being dominant at low temperatures. The tunneling mechanism may account for the weak temperature dependence of the times of impurity sorption at the lowest temperatures of the experiment.

### References

1. D.D.L. Chung, *J.Mater.Sci.* **37**, 1475 (2002).
2. C. Lee, X. Wei, J. W. Kysar, J. Hone, *Science* **321**, 385 (2008).
3. A.A. Balandin, S. Ghosh, W. Bao, I. Calizo, D. Teweldebrhan, F. Miao, C.N. Lau, *Nano Lett*. **8**, 902 (2008).
4. X. Huang, X. Qi, F. Boey, H. Zhang, *Chem. Soc. Rev.* **41**, 666 (2012).
5. F. Guinea, A. H. Castro Neto, N. M. R. Peres, *Phys. Rev. B* **73**, 245426 (2006).
6. K.S. Novoselov, Z. Jiang, Y. Zhang, S.V. Morozov, H.L. Stormer, U. Zeitler, J. C. Maan, G. S. Boebinger, P. Kim, A. K. Geim, *Science* **315**, 1379 (2007).
7. K.I. Bolotin, K.J. Sikes, Z. Jiang, M. Klima, G. Fudenberg, J. Hone, P. Kim, H.L. Stormer, *Solid State Commun.* **146,** 351 (2008).
8. J.S. Bunch, A.M. van der Zande, S.S. Verbridge, I.W. Frank, D.M. Tanenbaum, J.M. Parpia, H.G. Craighead, P.L. McEuen, *Science*, **315**, 490 (2007).
9. S. Hadlington, *Chem. World* **4**, 29 (2007).

10. B. Huang, Z. Y. Li, Z. R. Liu, G. Zhou, S. G. Hao, J. Wu, B. L. Gu, W. H. Duan, *J. Phys. Chem. C*, **112**, 13442 (2008).

11. B. G. Choi, H. Park, M. H. Yang, Y. M. Jung, J. Y. Park, S. Y. Lee, W. H. Hong and T. J. Park, *Nanoscale*, **2**, 2692 (2010).

12. K.J. Huang, D.J. Niu, J.Y. Sun, C.H. Han, Z.W. Wu, Y.L. Li, X.Q. Xiong, *Colloids Surf. B*, **82**, 543 (2011).

13. W.W. Hummers, Jr., R.E. Offeman, *J. Am. Chem. Soc.* **80,** 1339 (1958).

14. S. Mao, H. Pu and J. Chen, *RSC Advances* **2**, 2643, (2012).

15. A. B. Bourlinos, D. Gournis, D. Petridis, T. Szabo, A. Szeri, I. Dekany, *Langmuir*, **19**, 6050 (2003).

16. M. Hirata, T. Gotou, S. Horiuchi, M. Fujiwara, M. Ohba, *Carbon*, **42**, 2929 (2004).

17. H.K. He, C. Gao, *Science China Chemistry*, **54**, 397 (2011).

18. A. Grinou, Y.S. Yun, S.Y. Cho, H.H. Park, H.-J. Jin, *Materials* **5**, 2927 (2012).

19. D.R. Dreyer, S. Park, C.W. Bielawski, R.S. Ruoff, *Chem. Soc. Rev.* **39**, 228 (2010).

20. V. Singh, D. Joung, L. Zhai, S. Das, S.I. Khondaker, S. Seal, *Progress in Materials Science* **56**, 1178 (2011).

21. H.Q. Bao, Y.Z. Pan, Y. Ping, N.G. Sahoo, T.F. Wu, L. Li, J. Li, L.H. Gan, *Small*, **7,** 1569 (2011).

22. O. Akhavan, E. Ghaderi, S. Aghayee, Y. Fereydooni, A. Talebi, *J.Mater.Chem.* **22**, 13773 (2012).

23. J. Zhang, H. Yang, G. Shen, P. Cheng, J. Zhang, S. Guo, *Chem Commun*. **46**, 112 (2010).

24. J. Gao, F. Lui, Y. Lui, N. Ma, Z. Wang, X. Zhang, *Chem Mater.* **22**, 2213 (2010).

25. G. Srinivas, J. Burres, T. Yildirim, *Energy Environ. Sci.* **5**, 6453 (2012).

26. G. Srinivas, J. W. Burress, J. Ford, T. Yildirim, *J. Mater. Chem.* **21**, 11323 (2011).

27. L. Firlej and B. Kuchta, *Colloids and Surfaces A: Physicochem. Eng. Aspects* **241**, 149 (2004).

28. M.A. Strzhemechny, I.V. Legchenkova, *Fiz. Nizk. Temp.* **37**, 688 (2011) [*Low Temp. Phys.* **37**, 547 (2011)].

29. A.V. Dolbin, V.B. Esel'son, V.G. Gavrilko, V.G. Manzhelii, N.A. Vinnikov, S.N. Popov, *JETP Letters* **93**, 577 (2011).

30. A.V. Dolbin, V.B. Esel'son, V.G. Gavrilko, V.G. Manzhelii, N.A. Vinnikov, S.N. Popov, *Fiz. Nizk. Temp.* **38**, 1216 (2012) [Low Temp. Phys. **38**, 962 (2012)].

31. A.V. Dolbin, V. B. Esel'son, V. G. Gavrilko, V. G. Manzhelii, N. A. Vinnikov, R. M. Basnukaeva, I.I. Yaskovets, I.Yu. Uvarova, B. A. Danilchenko, *Fiz. Nizk. Temp.* (to be published)

32. A.V. Dolbin, V. B. Esel'son, V. G. Gavrilko, V. G. Manzhelii, N. A. Vinnikov, S. N. Popov and B. Sundqvist, *Fiz. Nizk. Temp.* **36**, 797 (2010) [*Low Temp. Phys.* **36**, 635 (2010)].
33. A.V. Dolbin, V. B. Esel'son, V. G. Gavrilko, V. G. Manzhelii, N. A. Vinnikov, S. N. Popov, and B. Sundqvist, *Fiz. Nizk. Temp.* **37**, 685 (2011) [*Low Temp. Phys.* **37**, 544 (2011)].
34. A. V. Dolbin, V. B. Esel'son, V. G. Gavrilko, V. G. Manzhelii, N. A. Vinnikov and S. N. Popov, *Fiz. Nizk. Temp.* **36**, 1352 (2010) [*Low Temp. Phys*. **36**, 1091 (2010)].
35. C. Vallés, J.D. Núñez, A.M. Benito, W.K. Maser, *Carbon* **50,** 835 (2012).
36. S. Stankovich, R.D. Piner, S.T. Nguyen, R.S. Ruoff, *Carbon* **44**, 3342 (2006).
37. V.C. Tung, M.J. Allen, Y. Yang, R.B. Kaner, *Nat. Nanotechnol.* **4**, 25 (2009).
38. L. J.Cote,; F.Kim, J.X. Huang, *J. Am. Chem. Soc.*, **131**, 1043 (2009).
39. S. Stankovich, D.A. Dikin, R.D. Piner, K.A. Kohlhaas, A. Kleinhammes, Y.Y. Jia, Y. Wu, S.T. Nguyen, R.S. Ruoff, *Carbon*, **45**, 1558 (2007).
40. C. Mattevi, G. Eda, S. Agnoli, S. Miller, K. A. Mkhoyan, O. Celik, D. Mostrogiovanni, G. Granozzi, E. Garfunkel and M. Chhowalla, *Adv. Funct. Mater.* **19**, 2577 (2009).
41. Q. Su, S. Pang, V. Alijani, C. Li, X. Feng, K. Müllen, *Adv. Mater.* **21**, 3191 (2009).
42. M.X. McAllister, J.L. Li, D.H. Adamson, H.C. Schniepp, A.A. Abdala, J. Liu, M. Herrera-Alonso, D.L. Milius, R. Car, R.K. Prud'homme, I.A. Aksay, *Chem. Mater.* **19**, 4396 (2007).
43. V. Tozzini, V. Pellegrini, *Phys. Chem. Chem. Phys.* **15**, 80 (2013).
44. G. Srinivas, Y. Zhu, R. Piner, N. Skipper, M. Ellerby, R. Ruoff, *Carbon*, **48**, 630 (2010).
45. A. Ghosh, K.S. Subrahmanyam, K.S. Krishna, S. Datta, A. Govindaraj, S.K. Pati, C.N.R. Rao, *J. Phys. Chem. C* **112**, 15704 (2008).
46. V.G. Manzhelii, A.I. Prokhvatilov, V.G. Gavrilko, A.P. Isakina. Structure and Thermodynamic Properties of Cryocrystals, *Begell House*, 316 (1999).
47. F. Costanzo, P.L. Silvestrelli, F. Ancilotto, *J. Chem. Theory Comput.* **8**, 1288 (2012).
48. Y.-K. Kwon, *Journal of the Korean Physical Society* **57**, 778 (2010).